\documentclass[12pt]{article}
\usepackage{graphicx}
\usepackage{amsmath}
\usepackage{amssymb}
\usepackage{caption2}
\begin{document}
\bibliographystyle{prsty}
\begin{center}
{\large {\bf \sc{   X(1835) as a baryonium state with QCD sum rules }}} \\[2mm]
Zhi-Gang Wang$^{1}$ \footnote{Corresponding author; E-mail,wangzgyiti@yahoo.com.cn.  } and Shao-Long Wan$^{2}$      \\
$^{1}$ Department of Physics, North China Electric Power University, Baoding 071003, P. R. China \\
$^{2}$ Department of Modern Physics, University of Science and Technology of China, Hefei 230026, P. R. China \\
\end{center}

\begin{abstract}
In this article, we take the point of view that the $X(1835)$ be a
baryonium state and calculate its mass within the framework of the
QCD sum rules approach. The numerical value of the mass of the
$X(1835)$ is consistent with the experimental data. There  may be
some baryonium component in the $X(1835)$ state.
\end{abstract}



\section{Introduction}
In 2003, the BES collaboration  observed a significant narrow
near-threshold enhancement in the proton-antiproton ($p\bar{p}$)
invariant mass spectrum  from the radiative decay $J/\psi\to\gamma
p\overline{p}$ \cite{BES03}.
 The enhancement can be fitted  with either an $S$- or $P$-wave Breit-Wigner
resonance function. In the case of the $S$-wave fitted form, the
Breit-Wigner mass $M = 1859 {}^{+3}_{-10}  {}^{+5}_{-25}  MeV$ and
the width $\Gamma < 30 MeV$. Recently the BES collaboration observed
a resonance  state X(1835) in the $\eta^{\prime}\pi^+\pi^-$
invariant mass spectrum in the process
$J/\psi\to\gamma\pi^{+}\pi^{-}\eta^{\prime}$  with the Breit-Wigner
mass $M=(1833.7\pm 6.2\pm 2.7)MeV$  and  the width $\Gamma=(67.7\pm
20.3\pm 7.7)MeV$ \cite{BES05}. Many theoretical works were
stimulated to interpret the nature and underlying structures of the
new particle, there exist many possibilities, for example, the
$p\bar{p}$ bound state, the pseudoscalar glueball, non-exotic state,
etc \cite{GluonballX,2Baryon,Huang06,TheoreticalWork}.

 The $X(1835)$ may be pseudoscalar glueball \cite{GluonballX}, which can take into account
   the observation  of the $X(1835)$ in the  $\eta' \pi^+\pi^-$ channel not in the $ \pi^0 \pi^+\pi^-$
 channel,  while  the strong coupling
 to the $p\bar{p}$  state  can  be related to the
large contribution of the gluon axial anomaly to the proton spin.
 The calculations with  lattice  QCD and QCD sum rules indicate that
  the pure scalar glueballs lie around $(1.5-1.7)GeV$
 and the pure pseudoscalar glueballs lie around $2.6 GeV$
\cite{LattM,Gluonball},  we have to resort to special mechanism to
pull the mass down to $1.835 GeV$.

The $X(1835)$ may also not be exotic state and be the candidate for
the second radial excitation of the $\eta'$ meson,  which fulfil the
pseudoscalar nonet $\pi(1800)$, $K(1830)$, $\eta(1760)$, $X(1835)$
\cite{Huang06}, such assignment can explain the mass, total decay
width, production rate and decay pattern phenomenologically. The
decay  $X(1835)\to \eta'\pi^+\pi^-$ takes place through the emission
of a pair of $S$-wave $\pi$ mesons, while the decay $X(1835)\to
\eta\pi\pi$ has not been observed experimentally  yet. Whether or
not there exists this decay mode is of great importance, further
experiments are needed to prove or exclude the possibility.

In this article, we take the point of view that the $X(1835)$  be a
baryonium with the quantum numbers $J^{PC}=0^{-+}$ \cite{2Baryon},
and calculate its mass in the framework of the QCD sum rules
approach \cite{SVZ79}. The  radiative decay of the $J/\psi$  is
generally believed to be glue-rich, which can explain the branching
ratio of the decay $J/\psi \to \gamma\eta^\prime$ (through
$J/\psi\to\gamma G\tilde G\to \gamma \eta'$) is large while the
decay  ratio of the $J/\psi \to \gamma\eta$ (through the $
\eta'-\eta$ mixing) is small, about $(9.8\pm 1.0) \times10^{-4}$
\cite{PDG06}. The observation of the $X(1835)$ in the $\eta'$
channel not in the $\eta$ channel may be due to the intermediate
virtual gluons are flavor-neutral and the $\eta'$ meson is mainly a
$SU(3)$ flavor singlet. The threshold $2m_p=1876>1835$ and the width
$\Gamma=68$, the decay $X(1835)\to p\bar{p}$ takes place through the
fall apart mechanism with re-arrangement in the color space, while
suppressed
 kinematically  and the decay occurs only through  the tail of the mass distribution.

 On the other hand, whether or not there exist some quark configurations which can result in the baryonium state
 is of great
importance   itself, we explore this possibility and propose a
special quark configuration, later experimental data can confirm or
reject this assumption.

The article is arranged as follows:  we derive the QCD sum rules for
the mass $M_X$ of the $X(1835)$  in section II; in section III,
numerical results and discussions; section VI is reserved for
conclusion.

\section{QCD sum rules for the $X(1835)$}
In the following, we write down  the two-point correlation function
$\Pi(q^2)$ in the framework of the QCD sum rules approach,
\begin{eqnarray}
\Pi(q^2)&=&i\int d^4x e^{iq \cdot x} \langle
0|T\{J_5(x)J_5^+(0)\}|0\rangle \, ,  \\
J_5(x)&=&\bar{J}_{kl}(x)i\gamma_5 J_{kl}(x) \, ,\\
J_{kl}(x)&=&\epsilon_{kmn} u^T_m(x)C\gamma^\alpha u_n(x)\gamma_5
\gamma_\alpha d_l(x) \, , \\
\lambda_X&=&\langle 0|J_5(0)|X\rangle \, .
\end{eqnarray}
Here the  $k$, $l$, $m$, $n$ are color indexes, the $C$ is charge
conjunction matrix, the  $\alpha$ is Lorentz index. The hexaquark
states  can be classified as "baryonia" if they can be described as
a single multiquark cluster, or "molecules" if they consist of
weakly bound $N\bar N$ pairs. We take the pseudoscalar
proton-antiproton type interpolating current $J_5(x)$ to represent
the $X(1835)$, if we smear the color indexes, the colored
constituent $J_{kl}(x)$ has the same
 structure as the Ioffe current $\eta(x)$ which interpolates the
proton,
\begin{eqnarray}
\eta(x)&=& \epsilon_{kmn} u^T_m(x)C\gamma^\alpha u_n(x)\gamma_5
\gamma_\alpha d_k(x),  \nonumber
 \end{eqnarray}
the color indexes $k$ and $l$ in the $J_{kl}(x)$ (in other words,
the strong color interactions) bind the two constituents as a single
multiquark cluster $uud \bar{u}\bar{u}\bar{d}$. If we take the color
singlet operator $\eta(x)$ as the basic constituent and choose
$\eta_5(x)=\bar{\eta}(x)i\gamma_5 \eta(x)$, the current $\eta_5(x)$
can interpolate a hexaquark state, whether the compact state or the
loosely deuteron-like $p\bar{p}$ bound  state, it is difficult to
separate the contributions of the bound state from the  scattering
$p\bar{p}$ state. In this article, we take the $X(1835)$ as a
baryonium state and choose the current $J_5(x)$, although the
$\eta_5(x)$  has non-vanishing coupling with the $X(1835)$.

 According to the basic assumption of current-hadron duality in
the QCD sum rules approach \cite{SVZ79}, we insert  a complete
series of intermediate states satisfying the unitarity   principle
with the same quantum numbers as the current operator $J_5(x)$
 into the correlation function in
Eq.(1)  to obtain the hadronic representation. After isolating the
pole term  of the lowest $X(1835)$  state, we obtain the following
result,
\begin{eqnarray}
\Pi(q^2)&=&\frac{\lambda_X^2}{M_X^2-q^2}+\cdots .
\end{eqnarray}

In the following, we briefly outline the calculations of the
operator product expansion in the deep Euclidean space.   In order
to evaluate the correlation function $\Pi(q^2)$ at the level of
quark-gluon degrees of freedom, we determine the quark propagator in
the presence of the quark and gluon condensates firstly\footnote{For
the $u$, $d$ and $s$ quarks, the current masses are small, it is
convenient to work  in the $x$-representation and adopt the external
field method, we follow the routine presented in page-28 and page-36
in the last (review) article of Ref.\cite{SVZ79} to carry out the
operator product expansion. For the technical details, one can
consult the excellent review "Hadron Properties from QCD Sum Rules",
L. J. Reinders, H. Rubinstein and S. Yazaki, Phys. Rept. {\bf 127}
(1985) 1. },
\begin{eqnarray}
S_{ab}(x)&\equiv&\langle0| T\{q_a(x)\bar{q}_b(0)\}|0 \rangle
\nonumber \\
&=&\frac{i\delta_{ab}\hat{x}}{2\pi^2x^4}
-\frac{ig_sG^{\mu\nu}_{ab}}{32\pi^2x^2}(\sigma_{\mu\nu}\hat{x}+\hat{x}\sigma_{\mu\nu})
-\frac{\delta_{ab}\langle\bar{q}q\rangle}{12}
-\frac{\delta_{ab}\langle g_s\bar{q}\sigma Gq\rangle
x^2}{192}+\cdots \, ,
\end{eqnarray}
where   the small masses of the $u$ and $d$ quarks are neglected.
Then we substitute the quark propagator into the following
correlation function to obtain the spectral density with the vacuum
condensates adding up to dimension-12,

\begin{eqnarray}
\Pi(q^2)&=&4i\int d^4x e^{iq\cdot x} \epsilon_{kij}
\epsilon_{kmn}\epsilon_{k'i'j'} \epsilon_{k'm'n'} \mbox{Tr} \left[
\gamma_\alpha \gamma_5 \gamma_\beta S_{ll'} (x)
\gamma_{\alpha'}\gamma_5 \gamma_{\beta'}S_{l'l}(-x)\right] \nonumber
\\
&& \mbox{Tr} \left[ C \gamma^{\beta'}S_{n'i}(-x)\gamma^\alpha C
S_{m'j}^T(-x) \right] \mbox{Tr} \left[ C
\gamma^{\beta}S_{ni'}(x)\gamma^{\alpha'} C S_{mj'}^T(x) \right] \, .
\end{eqnarray}
In Eq.(7), we have taken the assumption of the vacuum saturation for
the condensates, the high dimension vacuum condensates are always
 factorized to lower condensates with the vacuum saturation in the QCD sum rules,
 the factorization works well in the large $N_c$ limit. It is obvious that such an assumption
can not take into account some information  in the parameter space,
the straight forward calculations with the standard operator product
expansions can lead to a more general expression for the vacuum
condensates.  We take a simple routine in Eqs.(6-7) to simplify the
calculation and obtain the result in a special case, it is the
common approach to deal with the multiquark states with the QCD sum
rules, for example, the tetraquark states in Ref.\cite{Zhu06}. In
this article, we take into account the contributions from the quark
condensates $\langle \bar{q}q \rangle$, gluon condensates $\langle
\frac{\alpha_s GG}{\pi} \rangle$, mixed condensates $\langle
\bar{q}g_s \sigma G{q} \rangle $, and neglect the contributions from
other high dimension condensates which are suppressed by large
denominators and would not play significant roles. Once the
analytical results are obtained,
  then we can take the current-hadron duality  below the threshold
$s_0$ and perform the Borel transformation with respect to the
variable $Q^2=-q^2$, finally we obtain  the following sum rule,
\begin{eqnarray}
 \lambda_X^2 e^{-\frac{M_X^2}{M_B^2}}&=&\int_0^{s_0}ds
e^{-\frac{s}{M_B^2}}\frac{\mbox{Im}\Pi(s)}{\pi} \, ,
\end{eqnarray}

\begin{eqnarray}
\frac{\mbox{Im}\Pi(s)}{\pi}&=&
\frac{9s^7}{2^97!7!\pi^{10}}+\frac{s^4\langle\bar{q}q\rangle^2}{2^44!4!\pi^6}
+\frac{5s\langle\bar{q}q\rangle^4}{3^2
2\pi^2}+\frac{5s^3\langle\bar{q}q\rangle \langle\bar{q}g_s\sigma G
q\rangle }{2^9 3^2\pi^6}-
 \nonumber \\
&& \frac{5s^2  \langle\bar{q}g_s\sigma G q\rangle^2 }{2^{10} 3
\pi^6} -\frac{s^2\langle\bar{q}q\rangle^2
 }{2^6 3^3\pi^4}\langle\frac{\alpha_sGG}{\pi}\rangle-\frac{s\langle\bar{q}q\rangle\langle\bar{q}g_s\sigma G
 q\rangle
 }{2^8 3^2\pi^4}\langle\frac{\alpha_sGG}{\pi}\rangle \, .
\end{eqnarray}

Differentiate the above sum rule with respect to the variable
$\frac{1}{M_B^2}$, then eliminate the quantity $\lambda_X$, we
obtain the QCD sum rule for the mass,
\begin{eqnarray}
M_X^2&=&\int_0^{s_0}ds s
e^{-\frac{s}{M_B^2}}\frac{\mbox{Im}\Pi(s)}{\pi}   / \int_0^{s_0}ds
e^{-\frac{s}{M_B^2}}\frac{\mbox{Im}\Pi(s)}{\pi} \, .
\end{eqnarray}
It is easy to integrate over  the variable  $s$, we prefer this
formulation for simplicity. If we replace the $e^{-\frac{s}{M_B^2}}$
with $s^n$, $n=0,1,2,3, \dots$, we obtain the finite energy sum rule
(FESR) \cite{FESR},
\begin{eqnarray}
\lambda_X^2 M_X^{2n}&=&\int_0^{s_0}ds s^n\frac{\mbox{Im}\Pi(s)}{\pi}\, , \\
M_X^2&=&\int_0^{s_0}ds s^{n+1}\frac{\mbox{Im}\Pi(s)}{\pi}
/\int_0^{s_0}ds s^n\frac{\mbox{Im}\Pi(s)}{\pi} \,
 .
 \end{eqnarray}
 The threshold parameter $s_0$ is determined by the condition,
 \begin{eqnarray}
 \frac{d}{ds_0}M_X^2=0 \,.
 \end{eqnarray}
 The FESRs  correlate the ground state mass with the   continuum threshold $s_0$, and
separate the ground state from the  continuum contributions at the
very beginning, for  some pentaquark currents, there happen exist
reasonable stability regions  $s_0$ \cite{Narison04}. The weight
function $s^n$ enhances the continuum or the high mass resonances
rather than the lowest ground state, we must make sure that only the
lowest pole terms contribute to the FESR below the  $s_0$, in some
case, a naive stability region  $s_0$ can not guarantee    a
physically reasonable value of the $s_0$ \cite{Narison04}. For the
hexaquark state, the situation is much worse, the stability
condition in Eq.(13) can not be satisfied, we discard the FESR in
this article.
\section{Numerical results and discussions}
The input parameters are taken to be the standard values $\langle
\bar{u}u \rangle=\langle \bar{d}d \rangle=\langle \bar{q}q
\rangle=-(0.24\pm 0.01 GeV)^3$, $\langle \bar{q}g_s\sigma G q
\rangle=m_0^2\langle \bar{q}q \rangle$, $m_0^2=(0.8 \pm 0.1)GeV^2$,
$\langle \frac{\alpha_s GG}{\pi}\rangle=(0.33GeV)^4 $ and
$m_u=m_d=0$. In numerical calculation, we observe that the
contributions from the terms with the gluon condensate $\langle
\frac{\alpha_s GG}{\pi}\rangle $ are very small, and neglect the
uncertainty  of the gluon condensate. The main contributions to the
correlation function in Eq.(8) come from the terms with $\langle
\bar{q}q\rangle^2 $ and $\langle \bar{q}q\rangle \langle \bar{q}g_s
\sigma Gq\rangle $, about $85\%$; the contributions from the terms
with $\langle \bar{q}q\rangle^4$ and $\langle \bar{q}g_s \sigma G
q\rangle^2$ are about $35 \%$ and $20 \%$ respectively, and cancel
out with each other, the resulting
 net contributions are less than $15 \%$; the contribution comes  from
the perturbative term is very small, about $1\%$. We neglect the
contributions from other high dimension condensates which  are
suppressed by large denominators.
 In the QCD sum rules
with the interpolating currents constructed from the multiquark
configurations, the main contributions come from the terms with the
condensates   $\langle \bar{q}q \rangle$ and $\langle
\bar{s}s\rangle$ \cite{Oka94}, sometimes the mixed condensates
$\langle \bar{q}g_s\sigma G q \rangle$ and $\langle \bar{s}g_s\sigma
G s\rangle$ also play important roles \cite{MixC}.

In the following, we discuss the  criterion  for selecting the
threshold parameter $s_0$  and Borel parameter $M_B$  in the QCD sum
rules dealing with  the multiquark states.  For  the conventional
(two-quark) mesons and (three-quark) baryons, the hadronic  spectral
densities are experimentally well known, the separations between the
ground state and excited states are large enough, the "single-pole +
continuum states" model works well in representing the
phenomenological spectral densities. The continuum states can be
approximated by the contributions from the  asymptotic quarks and
gluons, and the single-pole dominance condition can be well
satisfied,
\begin{eqnarray}
\int_{s_0}^{\infty}\rho_{pert}e^{-\frac{s}{M_B^2}}ds <
\int^{s_0}_{0}(\rho_{pert}+\rho_{nonp})e^{-\frac{s}{M_B^2}}ds \, ,
\end{eqnarray}
 here the $\rho_{pert}$ and
$\rho_{nonp}$ stand for the contributions from the perturbative and
non-perturbative part of the spectral density respectively. From the
 condition in Eq.(14), we can obtain the maximal value of the Borel parameter
$M_B^{max}$, exceed this value, the single-pole dominance will be
spoiled. On the other hand, the Borel parameter must be chosen large
enough to warrant the convergence of the operator product expansion
and the contributions from the high dimension vacuum condensates
which are  known poorly are of minor importance, the minimal value
of the Borel parameter $M_B^{min}$ can be determined.

For the conventional  mesons and  baryons, the Borel window
$M_B^{max}-M_B^{min}$  is rather large and the reliable QCD sum
rules can be obtained. However, for the multiquark states i.e.
tetraquark states, pentaquark states, hexaquark states, etc, the
spectral densities $\rho\sim s^n$ with $n$ is larger than the
 ones for the conventional hadrons,   the integral
$\int_0^{\infty} s^n e^{-\frac{s}{M_B^2}} ds$ converges more slowly
\cite{Narison04}. If one do not want to release the condition   in
Eq.(14), we have to either postpone the threshold parameter $s_0$ to
very large values or choose very small values of the Borel parameter
$M_B^{max}$. With large values of the threshold parameter $s_0$ ,
for example, $s_0 \gg M_{gr}^2$, here the $gr$ stands for the ground
state, the contributions from the high resonance states and
continuum states are included in, we can not use the single-pole (or
ground state) approximation for the spectral densities; on the other
hand, with very small values of the Borel parameter $M_B^{max}$, the
operator product expansion is broken down, and the Borel window
$M_B^{max}-M_B^{min}$ shrinks to zero or negative values. We should
resort to the "multi-pole $+$ continuum states" to approximate the
phenomenological spectral densities. The onset of the continuum
states  is not abrupt,   the ground state, the first excited state,
the second excited state, etc, the continuum states appear
sequentially; the excited states may be loose  bound states and have
large widths. The threshold parameter $s_0$  is postponed to large
value, at that energy scale, the spectral densities can be well
approximated by the contributions from the asymptotic quarks and
gluons, and of minor importance for the sum rules.

The present experimental knowledge about the phenomenological
hadronic spectral densities of the multiquark states is  rather
vague, even the existence of the multiquark states is not confirmed
with confidence, and no knowledge about either there are high
resonances or not.

In this article,  the following criteria are taken. We choose the
suitable values of the Borel parameter $M_B$, on the one hand the
minimal values $M_B^{min}$  are large enough to warrant the
convergence of the operator product expansion, for
$M_B^{min}>\sqrt{3.5}GeV$, the dominating contributions come from
the terms with $\langle \bar{q}q\rangle^2 $ and $\langle
\bar{q}q\rangle \langle \bar{q}g_s \sigma Gq\rangle $, about $85\%$;
on the other hand the maximal values $M_B^{max}$   are small enough
to suppress the contributions from the high excited states and
continuum states,
 we choose the naive analysis  $e^{-s_0/(M_B^{max})^2}< e^{-1}$.
  For the hadronic spectral density, the more phenomenological
analysis is preferred, we approximate the spectral density with the
contribution from the single-pole term, the threshold parameter
$s_0$  is taken slightly above the ground state mass (
$\sqrt{s_0}>M_{gr}+\frac{\Gamma_{gr}}{2}$) to subtract the
contributions from the excited states and continuum states. In this
article, the threshold parameter $s_0$ is taken to be $\sqrt{s_0}=
(2.1-2.3)GeV
>1.9GeV$. It is reasonable for the Breit-Wigner mass $M=(1833.7\pm
6.2\pm 2.7)MeV$  and width $\Gamma=(67.7\pm 20.3\pm 7.7)MeV$. The
values of the $\lambda^2_X$ from the sum rules in Eq.(8) increase
quickly with $\sqrt{s_0}>2.3GeV$, which are shown in Fig.1, it may
serve as indication of the onset of the high resonances and
continuum states. From the Fig.1, we can see that the Borel
parameter can be chosen to be $M_B^2=(3.5-5.5)GeV^2$, $M_B^{max}
\leq \sqrt{s_0}\leq 2.3GeV$.

Finally, we obtain the value of the mass of the $X(1835)$,
\begin{eqnarray}
M_X&=&(1.9\pm0.1)GeV \, .
\end{eqnarray}
The numerical result is compatible with experimental data, one may
reject taking the value from the more phenomenological analysis as
quantitatively reliable, the result is qualitative at least. The
systematic studies with the random instanton liquid model indicate
that the masses of the diquarks are $m_S=(420\pm30) MeV$,
 $m_V=m_A=(940\pm20 )MeV$, $m_T=(570\pm 20) MeV$\cite{RILM}, here the $S$, $V$, $A$ and $T$ stand for the
 scalar, vector, axial-vector and tensor diquarks respectively.  In this article, the chosen    quark configuration
 has two diquark constituents,  a diquark ($\epsilon_{abc}u^T_b C \gamma_\mu u_c $)
  and an antidiquark ($\epsilon_{abc}\bar{u}_b C \gamma_\mu \bar{u}^T_c $), it is not surprise that the energy
  scale set by the diquarks is about $1.9GeV$.
  Different quark configurations can result in different masses for the hadrons, for
  example, the meson-meson type interpolating currents indicate the
  masses of the tetraquark states are less than $1GeV$ \cite{AZhang00}.

\begin{figure}
 \centering
 \includegraphics[totalheight=7cm,width=7cm]{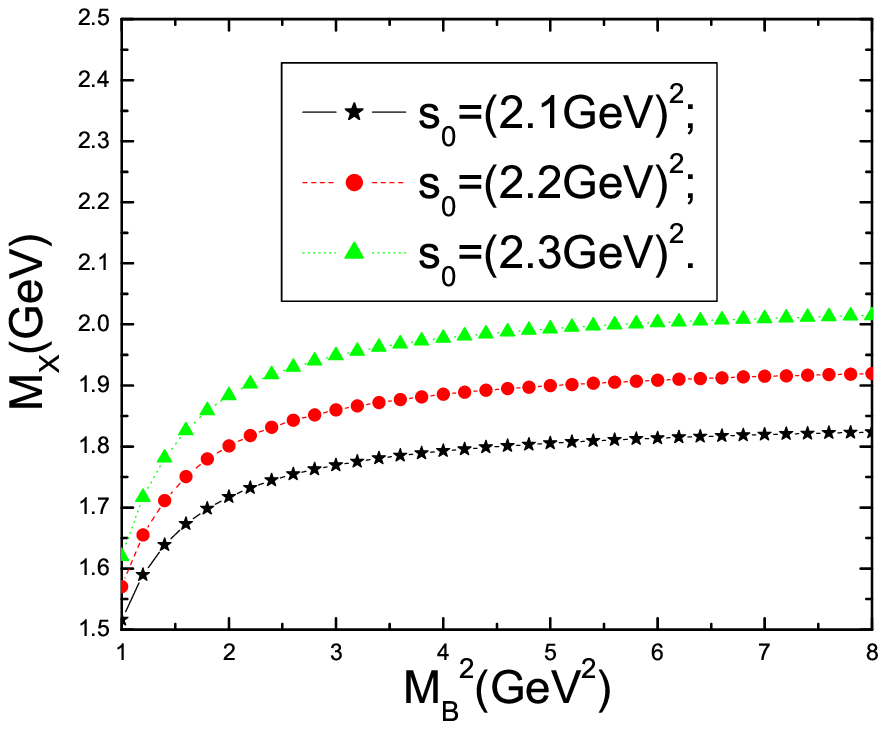}
  \includegraphics[totalheight=7cm,width=7cm]{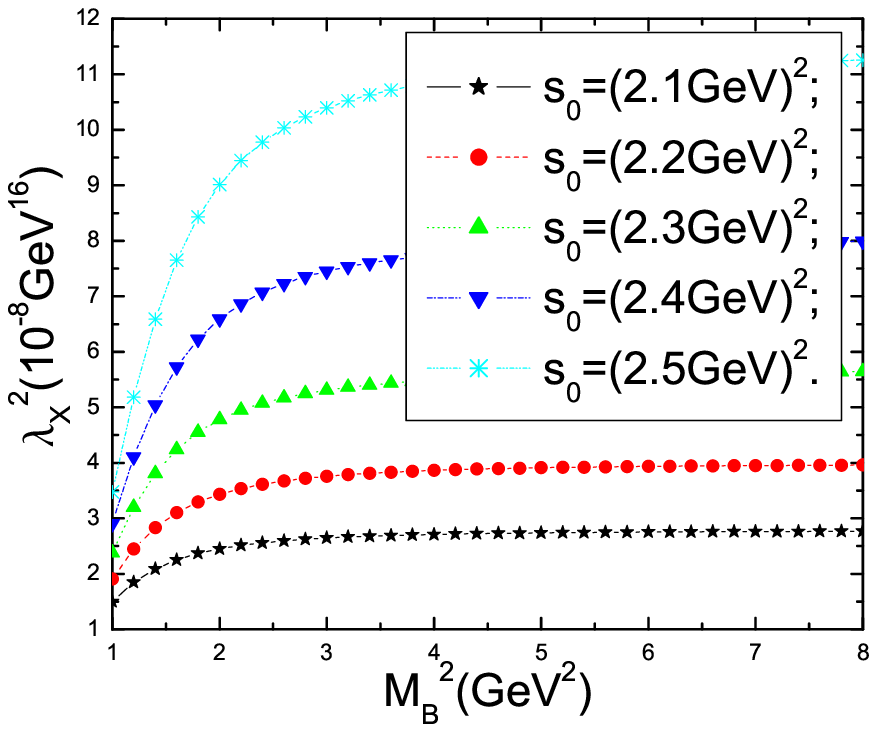}
 \caption{The   $M_X$ and $\lambda^2_X$ with the Borel parameter $M_B^2$ for $ \langle \bar{q}q \rangle=-(0.24GeV)^3$ and $m_0^2=0.8GeV^2$. }
\end{figure}

\section{Conclusion}
In this article, we take the point of view that the $X(1835)$ be a
baryonium state and calculate its mass within the framework of the
QCD sum rule approach. The numerical value of the mass of the
$X(1835)$ is consistent with the experimental data. There may be
some baryonium component in the $X(1835)$ state.

\section*{Acknowledgments}
This  work is supported by National Natural Science Foundation,
Grant Number 10405009,  and Key Program Foundation of NCEPU. The
authors are indebted to Dr. J.He (IHEP), Dr. X.B.Huang (PKU) and Dr.
L.Li (GSCAS) for numerous help, without them, the work would not be
finished.


\begin{thebibliography}{99}
\bibitem{BES03} J. Z. Bai et al,   Phys. Rev. Lett.
{\bf 91} (2003) 022001 .
\bibitem{BES05} M. Ablikim et al,
Phys. Rev. Lett. {\bf 95} (2005) 262001.


\bibitem{GluonballX} N. Kochelev and D. P. Min, Phys. Rev. {\bf D72} (2005)
    097502; G. Hao, C. F. Qiao and A. L. Zhang,
    hep-ph/0512214; X. G. He, X. Q. Li, X. Liu and J. P. Ma,
    hep-ph/0509140;  N. Kochelev and  D. P. Min,  Phys. Lett. {\bf B633} (2006)
    283.

\bibitem{2Baryon} A. Datta and P. J. O'Donnell,  Phys. Lett.  {\bf B567} (2003)
    273;  S. L. Zhu  and C. S. Gao, hep-ph/0507050; M. L. Yan, S. Li, B. Wu and B. Q. Ma,
    Phys. Rev. {\bf D72} (2005) 034027 ; G. J. Ding and M. L. Yan,
    hep-ph/0511186; G. J. Ding, J. l. Ping, M. L. Yan,
    hep-ph/0510013; G. J. Ding and M. L. Yan, Phys. Rev. {\bf C72} (2005)
    015208.

\bibitem{Huang06} T. Huang and S. L. Zhu, Phys. Rev. {\bf D73} (2006) 014023.

\bibitem{TheoreticalWork}
J. L. Rosner,  Phys. Rev.  {\bf D68} (2003) 014004;  D. V. Bugg,
Phys. Lett.  {\bf B598} (2004) 8;  B. S. Zou and H. C. Chiang, Phys.
Rev. {\bf D69} (2004) 034004;  B. Kerbikov, A. Stavinsky and V.
Fedotov, Phys. Rev.  {\bf C69} (2004)  055205;  X. A. Liu, X. Q.
Zeng, Y. B. Ding, X. Q. Li, H. Shen and P. N. Shen,
    hep-ph/0406118; B. Loiseau and  S.
Wycech, Int. J. Mod. Phys. {\bf A20} (2005) 1990; B. Loiseau and S.
Wycech, Phys. Rev. {\bf C72} (2005) 011001; X. G. He, X. Q. Li and
J. P. Ma, Phys. Rev.  {\bf D71} (2005) 014031.



\bibitem{LattM} C. Michael, hep-lat/0302001; and references therein.
\bibitem{Gluonball} K. Senba, M. Tanimoto, Phys. Lett. {\bf B105}
(1981) 297; S. Narison, Nucl. Phys. {\bf B509} (1998) 312; A. l.
Zhang, T. G. Steele, Nucl. Phys. {\bf A728} (2003) 165.





\bibitem{SVZ79}  M. A. Shifman, A. I. and Vainshtein and V. I. Zakharov,
Nucl. Phys. {\bf B147} (1979) 385, 448; L. J. Reinders, H.
Rubinstein and S. Yazaki, Phys. Rept. {\bf 127} (1985) 1.

\bibitem{PDG06} W. M. Yao et al, J. Phys. {\bf G33} (2006) 1.

\bibitem{Zhu06} H. X. Chen, A. Hosaka, S. L. Zhu, Phys. Rev. {\bf D74} (2006)
054001.

\bibitem{FESR}  S. Narison, QCD Spectral Sum Rules, World Scientific Lecture
Notes in Physics {\bf 26}; and references there in.

\bibitem{Narison04}  R. D. Matheus and S. Narison, hep-ph/0412063;
 W. Wei, P. Z. Huang, H. X. Chen, S. L. Zhu,
 JHEP {\bf 0507} (2005) 015;  Z. G. Wang, S. L. Wan and W. M. Yang,
 Eur. Phys. J. {\bf C45} (2006) 201.


\bibitem{Oka94}  N. Kodama, M. Oka and T. Hatsuda, Nucl. Phys. {\bf A580} (1994)
445;  Z. G. Wang, W. M. Yang and S. L. Wan, J. Phys. {\bf G31}
(2005) 971; Z. G. Wang and W. M. Yang, Eur. Phys. J. {\bf C42}
(2005) 89; Z. G. Wang, S. L. Wan, Nucl. Phys. {\bf A778} (2006) 22.

\bibitem{MixC} For example, B. L. Ioffe, A. G. Oganesian, JETP Lett. {\bf 80} (2004)
386; H. J. Lee, N. I. Kochelev, V. Vento, Phys. Rev. {\bf D73}
(2006) 014010.


\bibitem{RILM} T. Schafer, E. V. Shuryak, J. J. M. Verbaarschot, Nucl. Phys. {\bf B412} (1994)
143.


\bibitem{AZhang00} A. l. Zhang, Phys. Rev. {\bf D61} (2000) 114021.


\end{thebibliography}
\end{document}